\documentstyle[multicol,aps,psfig]{revtex}
\renewcommand{\narrowtext}{\begin{multicols}{2}
\global\columnwidth20.5pc\noindent}
\renewcommand{\widetext}{\end{multicols}
\global\columnwidth42.5pc}
\begin{document}
\draft
\preprint{18 September 1999}
\title{Simulated nuclear spin-lattice relaxation
       in Heisenberg ferrimagnets:\\
       Indirect observation of quadratic dispersion relations}
\author{Shoji Yamamoto}
\address{Department of Physics, Okayama University,
         Tsushima, Okayama 700-8530, Japan}
\date{26 October 1999}
\maketitle
\begin{abstract}
In response to recent proton spin relaxation-time measurements on
NiCu(pba)(H$_2$O)$_3$$\cdot$2H$_2$O with
$\mbox{pba}=1,3\mbox{-propylenebis(oxamato)}$, which is an excellent
one-dimensional ferrimagnetic Heisenberg model system of
spin-$(1,\frac{1}{2})$, we study the Raman relaxation process in
spin-$(S,s)$ quantum ferrimagnets on the assumption of predominantly
dipolar hyperfine interactions between protons and magnetic ions.
The relaxation time $T_1$ is formulated within the spin-wave theory
and is estimated as a function of temperature and an applied field
$H$ by a quantum Monte Carlo method.
The low-temperature behavior of the relaxation rate $T_1^{-1}$
qualitatively varies with $(S,s)$, while $T_1^{-1}$ is almost
proportional to $H^{-1/2}$ due to the characteristic dispersion
relations.
\end{abstract}
\narrowtext

   Quantum spin chains have been providing broadband topics and have
indeed invited numerous researchers to explore into them.
One of the hot topics is the mixed-spin system with two kinds of
antiferromagnetically exchange-coupled centers.
Besides the competition between the massive and massless states
\cite{Fuku09} which is an interesting issue in itself, quantum
ferrimagnetic phenomena have been attracting wide interest.
There already exists an accumulated chemical knowledge on bimetallic
chain compounds.
Kahn {\it et al.} \cite{Kahn95} succeeded in synthesizing the variety
of ferrimagnetic bimetallic chains such as
MCu(pba)(H$_2$O)$_3$$\cdot$$n$H$_2$O ($\mbox{M}=\mbox{Mn},\mbox{Ni}$)
\cite{Pei38} and MCu(pbaOH)(H$_2$O)$_3$$\cdot$$n$H$_2$O
($\mbox{M}=\mbox{Fe},\mbox{Co},\mbox{Ni}$) \cite{Koni25},
where $\mbox{pba}=1,3\mbox{-propylenebis(oxamato)}$ and
$\mbox{pbaOH}=2\mbox{-hydroxy-1,3-propylenebis(oxamato)}$.
Under close contact between chemists and physicists, not only
structural but also magnetic properties of them were elucidated
\cite{Dril13}.
It may be a few mathematical investigations
\cite{Alca67,Tian55,Shen80} that renewed our enthusiasm toward the
subject.
Motivated by these works, several authors carried out spin-wave
analyses \cite{Pati94,Breh21} on the alternating-spin Heisenberg
chains and pointed out the coexistence of
the ferromagnetic and antiferromagnetic branches in the low-lying
excitations.
Modified spin-wave theories \cite{Arov16,Taka33} were further applied
to the thermodynamics \cite{Yama08,Wu} and a
ferromagnetic-to-antiferromagnetic crossover was found in the
temperature dependences of the specific heat and the magnetic
susceptibility.
The dual structure of the excitations results in rich physics in a
magnetic field as well, {\it e.g.} the double-peak structure of the
specific heat \cite{Mais08} and quantized plateaux in the
ground-state magnetization curve \cite{Kura62,Saka}.
Quantum ferrimagnetism and related phenomena are still central
subjects in the field of material science.
A family of one-dimensional oxides, Sr$_3$MM$'$O$_6$
($\mbox{M}=\mbox{Ni},\mbox{Cu},\mbox{Zn};\
 \mbox{M}'=\mbox{Pt},\mbox{Ir}$),
consisting of alternating M$'$O$_6$ octahedra and MO$_6$ trigonal
prisms, visualizes the competition between various magnetic phases
\cite{Nguy00}, whereas an ordered double perovskite, Sr$_2$FeMoO$_6$,
acts as a half-metallic ferrimagnet \cite{Koba77,Kim37}, exhibiting a
novel magnetoresistive behavior.

\begin{figure}
\begin{flushleft}
\ \ \mbox{\psfig{figure=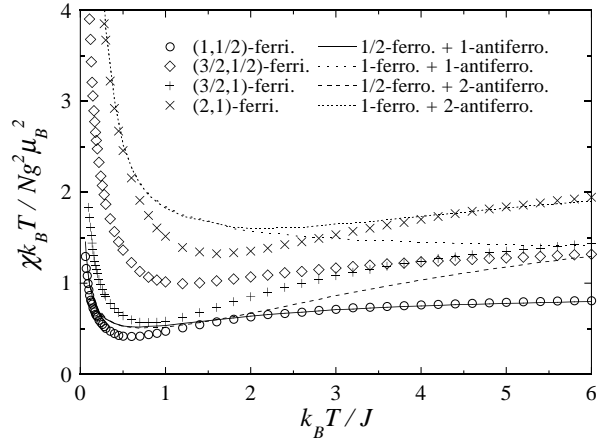,width=80mm,angle=0}}
\end{flushleft}
\caption{Temperature dependences of the magnetic susceptibility times
         temperature for the Heisenberg ferrimagnetic chains of
         $N=32$, compared with the sums of those for the
         spin-$(S-s)$ ferromagnetic and spin-$(2s)$ antiferromagnetic
         Heisenberg chains of $N$ spins.}
\label{F:chiT}
\end{figure}

   In such circumstances, the proton spin relaxation time $T_1$ has
recently been measured \cite{Fuji} for
NiCu(pba)(H$_2$O)$_3$$\cdot$2H$_2$O,
which is typical of the spin-$(1,\frac{1}{2})$ ferrimagnetic
Heisenberg chain, in an attempt to reveal the electron-spin dynamics
inherent in quantum ferrimagnets.
Although the susceptibility-temperature product, $\chi T$, shows a
minimum in its temperature dependence (see Fig. \ref{F:chiT} bellow),
the observed relaxation rate $T_1^{-1}$ is monotonically decreasing
with the increase of temperature.
The most interesting observation is the dependence of the relaxation
rate on the applied magnetic field $H$, which looks like
$T_1^{-1}\propto H^{-1/2}$, though the authors did not rule out a
logarithmic behavior, $T_1^{-1}\sim{\rm ln}H$, based on a rather
approximate argument.
We here present a rapid communication on our numerical simulation of
$T_1$ all the more because the measurements reported are highly
stimulative but still preliminary.
We show that {\it $T_1$ should indeed act as
$T_1^{-1}\propto H^{-1/2}$ for general spin-$(S,s)$ ferrimagnetic
Heisenberg chains due to the characteristic dispersion relations.}
The $H^{-1/2}$-dependence of $T_1^{-1}$ may first remind us of the
diffusive behavior of the spin correlation function \cite{Bouc98}.
However, a totally different mechanism may cause the field-dependent
relaxation rate in the present system.
Although the dynamic behavior \cite{Yama11} of quantum ferrimagnets
is a fascinating subject, measurements in this direction seem to be
still in their early stage.
We expect the present calculation to accelerate experimental
investigations into the dynamic properties.

   We consider two kinds of spins $S$ and $s$ ($S>s$) alternating on
a ring with antiferromagnetic exchange coupling between nearest
neighbors, as described by the Hamiltonian
\begin{equation}
   {\cal H}
      =J\sum_{j=1}^N
        \left(
         \mbox{\boldmath$S$}_{j} \cdot \mbox{\boldmath$s$}_{j}
        +\mbox{\boldmath$s$}_{j} \cdot \mbox{\boldmath$S$}_{j+1}
        \right)
      -g\mu_{\rm B} H\sum_{j=1}^N(S_j^z+s_j^z)\,,
   \label{E:H}
\end{equation}
where we have set the $g$ factors of spins $S$ and $s$ both equal to
$g$.
Keeping in mind ferrimagnetic bimetallic chain compounds such as
NiCu(pba)(H$_2$O)$_3$$\cdot$2H$_2$O \cite{Pei38}, we consider the
proton spin relaxation of the two-magnon (Raman) type \cite{Beem59},
which is usually the most dominant process due to the
energy-conservation requirement for the electronic-nuclear spin
system.
The relaxation rate of our interest, originating from longitudinal
spin fluctuations, is generally represented as
\widetext
\begin{equation}
   \frac{1}{T_1}
    =\frac{4\pi(g\mu_{\rm B}\hbar\gamma_{\rm N})^2}
          {\sum_n{\rm e}^{-E_n/k_{\rm B}T}}
     \sum_{n,m}{\rm e}^{-E_n/k_{\rm B}T}
     \big|
      \langle m|{\scriptstyle\sum_j}(A_j^zS_j^z+a_j^zs_j^z)|n\rangle
     \big|^2
     \,\delta(E_m-E_n-\hbar\omega_{\rm N})\,,
\label{E:T1def}
\end{equation}
\narrowtext
where
$A_j^z$ and $a_j^z$ are the coupling constants for the dipolar
hyperfine interactions between protons and electron spins,
$\omega_{\rm N}\equiv\gamma_{\rm N}H$ is the Larmor frequency of the
proton with $\gamma_{\rm N}$ being the gyromagnetic ratio, and the
summation $\sum_n$ is taken over all the eigenstates of $|n\rangle$
with energy $E_n$.
\vskip 2mm
\begin{figure}
\begin{flushleft}
\ \ \mbox{\psfig{figure=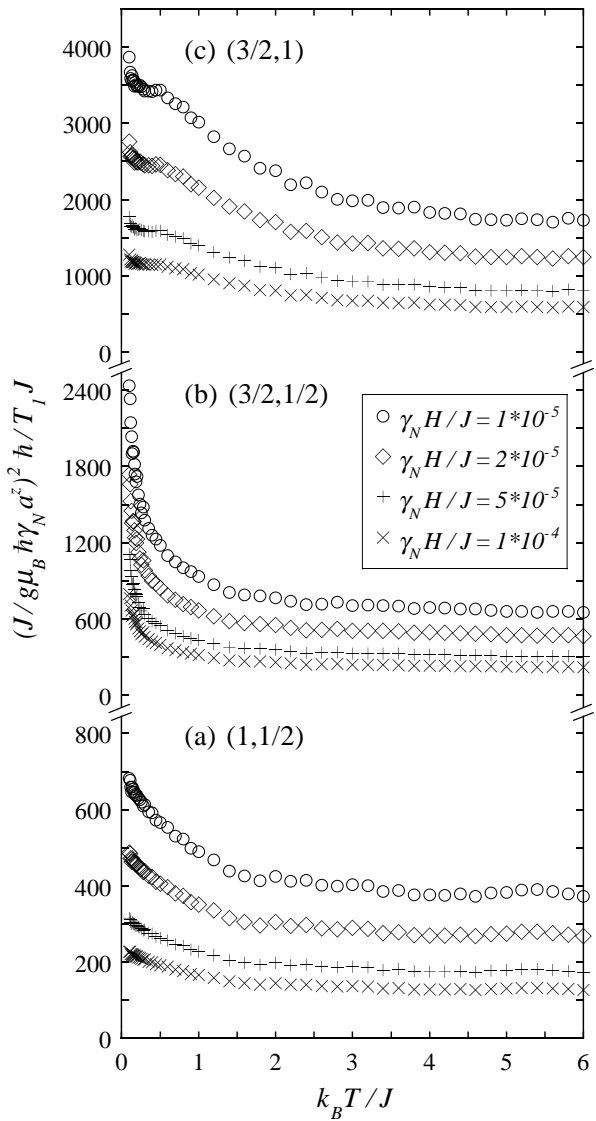,width=80mm,angle=0}}
\end{flushleft}
\caption{Temperature dependences of the proton spin relaxation rate
         at various values of the applied magnetic field:
         (a) $(S,s)=(1,\frac{1}{2})$,
         (b) $(S,s)=(\frac{3}{2},\frac{1}{2})$, and
         (c) $(S,s)=(\frac{3}{2},1)$.}
\label{F:T1T}
\end{figure}
\vskip 2mm

   Let us introduce the spin-deviation operators within the
linearized spin-wave theory \cite{Pati94,Breh21} via
$S_j^+=\sqrt{2S}\,a_j$,
$S_j^z=S-a_j^\dagger a_j$,
$s_j^+=\sqrt{2s}\,b_j^\dagger$, and
$s_j^z=-s+b_j^\dagger b_j$.
Then the diagonalized spin-wave Hamiltonian without constant terms is
represented as
\begin{equation}
   {\cal H}=
      \sum_k
      \left(
       \omega_{k}^-\alpha_k^\dagger\alpha_k
      +\omega_{k}^+\beta_k^\dagger \beta_k
      \right)\,,
   \label{E:SWH}
\end{equation}
where $\alpha_k$ and $\beta_k$ describe the ferromagnetic and
antiferromagnetic spin waves, respectively, and are related with the
sublattice bosons via
$\alpha_k=a_k{\rm cosh}\theta_k+b_k^\dagger{\rm sinh}\theta_k$ and
$\beta_k=a_k^\dagger{\rm sinh}\theta_k+b_k{\rm cosh}\theta_k$ with
$a_k=N^{-1/2}\sum_j{\rm e}^{ {\rm i}k(j-1/4)}a_j$,
$b_k=N^{-1/2}\sum_j{\rm e}^{-{\rm i}k(j+1/4)}b_j$, and
${\rm tanh}(2\theta_k)=2\sqrt{Ss}{\rm cos}(k/2)/(S+s)$.
Here twice the lattice constant has been taken as unity.
The dispersion relations are given by
\begin{equation}
   \omega_{k}^\pm=\omega_k\pm(S-s)J\mp g\mu_{\rm B}H\,,
   \label{E:DSPpm}
\end{equation}
with
\begin{equation}
   \omega_k=J\sqrt{(S-s)^2+4Ss\sin^2(k/2)}\,.
   \label{E:DSP}
\end{equation}
Now, having in mind the significant difference between the electronic
and nuclear energy scales ($\omega_{\rm N}\alt 10^{-5}J$), the
relaxation rate can be expressed in terms of the spin waves as
\begin{figure}
\begin{flushleft}
\ \ \mbox{\psfig{figure=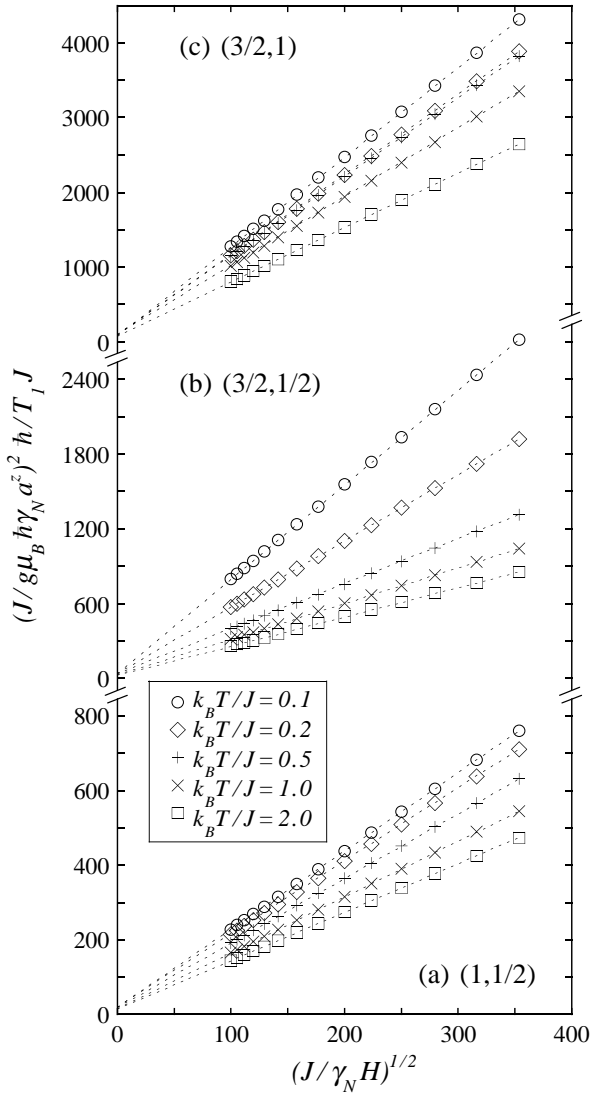,width=80mm,angle=0}}
\end{flushleft}
\caption{Field dependences of the proton spin relaxation rate
         at various values of temperature:
         (a) $(S,s)=(1,\frac{1}{2})$,
         (b) $(S,s)=(\frac{3}{2},\frac{1}{2})$, and
         (c) $(S,s)=(\frac{3}{2},1)$.}
\label{F:T1H}
\end{figure}
\vskip5mm
\begin{eqnarray}
   \frac{1}{T_1}
   =&&\frac{4\hbar}{NJ}(g\mu_{\rm B}\gamma_{\rm N})^2
      \sum_k\frac{S-s}{\sqrt{(Ssk)^2+2(S-s)Ss\hbar\omega_{\rm N}/J}}
   \nonumber \\
   &&\times
   \bigl[
    (A^z{\rm cosh}^2\theta_k-a^z{\rm sinh}^2\theta_k)^2
    n_k^-(n_k^- +1)
   \nonumber \\
   && \ 
   +(A^z{\rm sinh}^2\theta_k-a^z{\rm cosh}^2\theta_k)^2
    n_k^+(n_k^+ +1)
   \bigr]\,,
   \label{E:T1SW}
\end{eqnarray}
where
$A^z\equiv\sum_j A_j^z$ and $a^z\equiv\sum_j a_j^z$ are the $k=0$
Fourier components of the hyperfine coupling constants, and
$n_k^-\equiv\langle\alpha_k^\dagger\alpha_k\rangle$ and
$n_k^+\equiv\langle \beta_k^\dagger\beta_k \rangle$ are the thermal
averages of the numbers of the spin waves at a given temperature.
We then have an idea of evaluating $n_k^\pm$ by a quantum Monte Carlo
method.
Though the divergence of the ground-state magnetization, which
plagues the spin-wave treatment of low-dimensional antiferromagnets,
does not persist in our model, $n_k^\pm$ still diverge as temperature
increases as far as we naively estimate them with the noncompact
bosonic Hamiltonian (\ref{E:SWH}).
However, introducing a certain constraint, for instance, on the
magnetization, we can totally remove the difficulty and obtain a fine
description of the thermodynamics \cite{Yama08}, where it should be
stressed that the dispersion (\ref{E:DSPpm}) is still preserved.
Thus, relying upon the Bogoliubov transformation, we perform the
Monte Carlo sampling for the relevant spin operators
\begin{equation}
   \left.
   \begin{array}{l}
     \alpha_k
     ={\displaystyle\frac{{\rm cosh}\theta_k}{\sqrt{2N}}}
      {\displaystyle\sum_j}{\rm e}^{ {\rm i}k(j-\frac{1}{4})}S_j^+
     +{\displaystyle\frac{{\rm sinh}\theta_k}{\sqrt{ N}}}
      {\displaystyle\sum_j}{\rm e}^{ {\rm i}k(j+\frac{1}{4})}s_j^+
     \,,\\
     \beta_k
     ={\displaystyle\frac{{\rm sinh}\theta_k}{\sqrt{2N}}}
      {\displaystyle\sum_j}{\rm e}^{-{\rm i}k(j-\frac{1}{4})}S_j^-
     +{\displaystyle\frac{{\rm cosh}\theta_k}{\sqrt{ N}}}
      {\displaystyle\sum_j}{\rm e}^{-{\rm i}k(j+\frac{1}{4})}s_j^-
     \,,\\
   \end{array}
   \right.
\end{equation}
with the original compact Hamiltonian (\ref{E:H}).
While we base the relaxation process on the spin-wave excitations, we
take grand-canonical averages within the original system.
The thus-estimated $n_k^\pm$ are indeed consistent with the exact
calculations [23] of the dynamic structure factors
$S^{\pm}(k,\omega)=\sum_n
 \left\vert\langle n|S_k^\pm +s_k^\pm|{\rm g}\rangle\right\vert^2
 \delta(\omega-E_n+E_{\rm g})$
through the relation $n_k^\pm=\sum_\omega S^\pm(k,\omega)$, where
$|{\rm g}\rangle$ is the ground state with energy $E_{\rm g}$ and
magnetization $(S-s)N$.
Since the applied field $H$ is small enough in practice
($g\mu_{\rm B}H\alt 10^{-2}J$) \cite{Fuji}, we neglect the Zeeman
term of the Hamiltonian (\ref{E:H}) in the numerical treatment.
Considering the case of the contributing protons being located near
the smaller-spin magnetic ions \cite{Pei38}, we set $A^z/a^z$ equal
to $1/3$.

   We show in Fig. \ref{F:T1T} the relaxation rate as a function of
temperature.
The calculations in the case of $(S,s)=(1,\frac{1}{2})$ are
qualitatively consistent with the experimental observations
\cite{Fuji}, where $T_1^{-1}$ is a monotonically decreasing function
of temperature.
$T_1^{-1}$ hardly depends on temperature except for the
effectively-low-temperature region satisfying $k_{\rm B}T\alt W^-$,
where $W^-$ is the ferromagnetic band width, which may qualitatively
be identified with $\omega_{k=\pi}^- - \omega_{k=0}^-$.
Considering the dispersion relations (\ref{E:DSPpm}), it is convincing
that the ferromagnetic, rather than antiferromagnetic, nature
dominates the low-temperature relaxation process.
In this context, we show in Fig. \ref{F:chiT} the numerical
calculations of $\chi T$, which is closely related with $T_1^{-1}$
\cite{Hone65}.
$\chi T$ is a monotonically decreasing function in ferromagnets,
while a monotonically increasing function in antiferromagnets.
Thus we learn that ferrimagnets display the mixed behavior in their
temperature dependences of $\chi T$.
Recently it has been pointed out \cite{Yama24} that
Spin-$(S,s)$ ferrimagnetic chains behave similar to combinations of
spin-$(S-s)$ ferromagnetic and spin-$(2s)$ antiferromagnetic chains
{\it provided} $S=2s$, which is again illustrated in Fig.
\ref{F:chiT}.
For $S>2s$, quantum ferrimagnets are predominantly ferromagnetic
rather than antiferromagnetic, where $\chi T$ more rapidly decreases
than the, let us say, {\it balanced} behavior,
$(\chi^{(S-s){\scriptstyle\mbox{-}}{\rm ferro}}+
 \chi^{(2s){\scriptstyle\mbox{-}}{\rm antiferro}})T$,
whereas for $S<2s$, vice versa, where the low-temperature decrease is
duller than the {\it balanced} behavior.
It is quite interesting to observe Fig. \ref{F:T1T} from this point
of view.
The relatively rapid decrease of $T_1^{-1}$ at low temperatures for
$(S,s)=(\frac{3}{2},\frac{1}{2})$ must be of predominantly
ferromagnetic aspect, while the shoulder-like, even increasing,
behavior at low temperatures for $(S,s)=(\frac{3}{2},1)$ can be
regarded as an antiferromagnetic feature.
However, the overall gradual increase at high temperatures, as
observed for $\chi T$, does not appear.
It is the field-dependent prefactor to $n_k^\pm$ in Eq.
(\ref{E:T1SW}), coming from the energy-conservation requirement
$\delta(E_m-E_n-\omega_{\rm N})$ in Eq. (\ref{E:T1def}), that
suppresses the {\it latent} increasing behavior at high temperatures.
Since $\omega_{\rm N}\ll J$, the $k=0$ component is predominant in
the summation in Eq. (\ref{E:T1SW}).
In general, with the increase of temperature, $n_k^\pm$ decrease in
the vicinity of the zone center, whereas they increase near the zone
boundary \cite{Yama11}, where we note that $n_k^\pm$ both have their
peaks at $k=0$.
Therefore, the increase of the field, reducing the predominance of
the $k=0$ component, smears out the temperature dependence of
$T_1^{-1}$.

   The characteristic prefactor in Eq. (\ref{E:T1SW}) further leads
to the unique field dependence of $T_1^{-1}$.
In Fig. \ref{F:T1H} we plot $T_1^{-1}$ as a function of
$\omega_{\rm N}=\gamma_{\rm N}H$.
$T_1^{-1}$ is highly linear with respect to $H^{-1/2}$.
We stress that the observations in Fig. \ref{F:T1H} originate from
the quadratic dispersion relations peculiar to this system and
therefore the $H\rightarrow\infty$ extrapolation of $T_1^{-1}$
results in a considerably small value.
Thus the present phenomena are distinguishable from the diffusive
behavior,
$\langle S_q^+(t)S_{-q}^-(0)\rangle\propto{\rm e}^{-Dq^2t}$,
which also causes the field dependence of the form \cite{Bouc98}
\begin{equation}
   \frac{1}{T_1}=P+Q\sqrt{\frac{J}{\gamma_{\rm N}H}}\,.
   \label{E:T1FD}
\end{equation}
The constant $P$ is inherent in the fluctuation-dominant spin
dynamics.
For (CH$_3$)$_4$NMnCl$_3$ and LiV$_2$O$_5$, which are both well known
to be Heisenberg linear-chain compounds whose spin correlation
functions behave diffusively at long times, $P/Q$ was estimated to be
$139$ \cite{Bouc98} (in the high-temperature limit) and $376$
\cite{Fuji45} (at $k_{\rm B}T/J\simeq 0.65$), respectively.
On the other hand, the spin-$(1,\frac{1}{2})$ Heisenberg
ferrimagnetic chain compound NiCu(pba)(H$_2$O)$_3$$\cdot$2H$_2$O
exhibits $P/Q\simeq 12$ \cite{Fuji} (at $k_{\rm B}T/J\simeq 2.3$),
which is much smaller than those of the spin-diffusive materials but
is in good agreement with the present numerical findings
(Table \ref{T:T1P}).
Therefore {\it the $T_1$ measurements could implicitly demonstrate
the low-energy quadratic dispersion relations of quantum
ferrimagnets.}
When the $XY$-type exchange anisotropy moves the model away from the
Heisenberg point, the chain turns critical, showing a linear
dispersion \cite{Alca67,Ono76}, and the present field dependence of
$T_1^{-1}$ should no more be expected.
Thus the nuclear-magnetic-resonance, as well as neutron-scattering,
measurements allow us to reveal the low-energy structure efficiently.
We believe that the present calculations will greatly motivate
further experimental investigations into quantum ferrimagnets.

   The author is indebted to Dr. N. Fujiwara for helpful comments and
fruitful discussion on proton-NMR measurements for quantum
ferrimagnets.
This work was supported by the Japanese Ministry of Education,
Science, and Culture and by Sanyo-Broadcasting Foundation for Science
and Culture.
The numerical computation was done in part using the facility of the
Supercomputer Center, Institute for Solid State Physics, University of
Tokyo.

\begin{table}
\caption{$P/Q$ in the case of $T_1$ being fitted to
         $T_1^{-1}=P+Q(\gamma_{\rm N}H/J)^{-1/2}$.
         The estimates in the limit of
         $k_{\rm B}T/J\rightarrow\infty$ are obtained through a rough
         extrapolation and are thus no more than for reference.}
\begin{tabular}{ccccc}
$k_{\rm B}T/J$ & $(1,\frac{1}{2})$ & $(\frac{3}{2},\frac{1}{2})$ &
                 $(\frac{3}{2},1)$ & $(2,1)$ \\
\noalign{\vskip 1mm}
\tableline
\noalign{\vskip 1mm}
$0.5$    & $10.69$ & $ 9.35$ & $ 8.95$ & $ 8.17$ \\
$1.0$    & $11.10$ & $10.33$ & $10.23$ & $ 9.35$ \\
$2.0$    & $11.30$ & $10.49$ & $10.60$ & $10.28$ \\
$3.0$    & $11.30$ & $10.68$ & $11.49$ & $10.80$ \\
$5.0$    & $11.56$ & $10.93$ & $11.79$ & $11.24$ \\
$\infty$ & $11.6 $ & $11.0 $ & $12.2 $ & $11.8 $ \\
\end{tabular}
\label{T:T1P}
\end{table}

\widetext
\end{document}